\documentclass[referee,usenatbib]{iupartex}
\usepackage{newtxtext,newtxmath}
\usepackage[T1]{fontenc}
\usepackage{ae,aecompl}

\usepackage{multicol}   
\usepackage{bm}		    
\usepackage{pdflscape}	

\title[Trumpler 2]{Analysis of the young open cluster Trumpler 2 using Gaia DR3 data}

\author[Ta\c {s}demir \& Yontan]{%
S. Ta\c {s}demir$^{1\cc}$\orcid{0000-0003-1339-9148},
and
T. Yontan$^{2}$ \orcid{0000-0002-5657-6194}
\affsep \\
$^1$Istanbul University, Institute of Graduate Studies in Science, Programme of Astronomy and Space Sciences, 34116, Beyaz{\i}t, Istanbul, Turkey\\
$^2$Istanbul University, Faculty of Science, Department of Astronomy and Space Sciences, 34119, Beyaz\i t, Istanbul, Turkey\\
}

\corres{S. Ta\c {s}demir}{tasdemir.seval@ogr.iu.edu.tr}

\pubyear{20XX}

\doiheader{XXXXXXX/PAR.20XX.00000}
\date{
	\pSubmit{00.00.0000} 
	\pRevReq{00.00.0000}
	\pLastRevRec{00.00.0000}
	\pAccept{00.00.0000}
	\pPubOnl{00.00.0000}
}
\volume{0}
\volnumber{1}

\nolinenumbers 

\begin{document}
\label{firstpage}
\pagerange{\pageref*{firstpage}--\pageref*{lastpage}}
\maketitle

\begin{abstract}

We present an investigation of  the open cluster Trumpler 2 using {\it Gaia} DR3 photometric, astrometric and spectroscopic  data. 
92 stars were identified as likely members of the cluster, with membership probabilities greater than 0.5. The mean proper-motion components of the cluster are derived as ($\mu_{\alpha}\cos \delta$, $\mu_{\delta}$)=($1.494 \pm 0.004$, $-5.386 \pm 0.005$) mas yr$^{-1}$. By comparing the {\it Gaia} based colour-magnitude diagram with the {\sc PARSEC} isochrones scaled to {$z=0.0088$}, age, distance modulus and reddening are simultaneously estimated as $t=110 \pm 10$ Myr, {$\mu=10.027 \pm0.149$} mag and {$E(G_{\rm BP}-G_{\rm RP})=0.452\pm 0.019$} mag, respectively. The total mass of the cluster is estimated as 162 $M/M_{\odot}$ based on the stars with membership probabilities $P > 0$. The Mass function slope is derived to be $\Gamma = 1.33 \pm 0.13$ for Trumpler 2. This value is in a good agreement with that of of Salpeter. Galactic orbit analyses show that the Trumpler 2 orbits in a boxy pattern outside the solar circle and belongs to the young thin-disc component of the Galaxy. 
\end{abstract}

\begin{keywords}
Galaxy: open clusters and associations; individual: Trumpler 2,  
Galaxy: Stellar kinematics, stars: Hertzsprung Russell (HR) diagram,
\end{keywords}



\section{Introduction}
\label{sec:Introduction}
Open Clusters (OCs) are celestial objects consisting of stars bound to each under via mutual gravitational attraction. OCs are formed when a molecular cloud collapses. This means that the member stars of an OC formed under similar physical conditions and at a similar time. Hence, the stars that belong to the OCs have similar properties such as distance to Sun, metallicity and age. However stellar masses vary, leading to a range of stellar luminosities being observed in such clusters. These features make  OCs valuable and interesting laboratories for understanding the formation and evolution of our Galaxy \citep{Lada_2003}. 

The available information in the literature for the Trumpler 2 is as follows: Trumpler 2 ($\alpha=02^{\rm h} 36^{\rm m} 55^{\rm s}.7, \delta= +55^{\circ} 54^{\rm '} 18^{\rm''}$; $l=137^{\circ}.3863, b=-03^{\circ}.9778$, J2000) is located at the Perseus Constellation. Since the cluster has low central concentration, it is classified in the Trumpler catalogue as II 2p \citep{Trumpler_1930}. \citet{Frolov_2006} determined proper motion components  of Trumpler 2 on 6 plates with a maximal epoch difference of 63 years by considering the positions of about 3,000 stars within the $B$ $\sim < $ 16.25 magnitude. By taking into account 148  stars assessed as the most probable cluster members, based on astrometric and photometric criteria, \citet{Frolov_2006} obtained the colour-excess, distance modulus, distance and age as $0.40 \pm 0.06$ mag, $10.50\pm 0.35$ mag, 725 pc and 89 Myr, respectively. \citet{Zejda_2012} determined the mean proper-motion components of 879 open clusters including Trumpler 2,  based on 4 member stars in that cluster, as ($\mu_{\alpha}\cos\delta, \mu_{\delta})=(1.0~\pm~0.3, -4.6\pm 0.3$) mas yr$^{-1}$. \citet{Dias_2014} obtained the mean proper-motion components by using the U.S. Naval Observatory CCD Astrograph Catalogue \citep[UCAC4;][]{Zacharias_2013} as ($\mu_{\alpha}\cos\delta, \mu_{\delta})=(0.11 \pm 1.48, -3.49 \pm 2.23$) mas yr$^{-1}$. They also identified 346 stars as members of Trumpler 2. \citet{Paunzen_2014} studied chemically peculiar stars inside 10 open clusters including Trumpler 2. Moreover, it is confirmed that HD 16080 and BD+55~664 are members of Trumpler 2. As part of a survey of 1241 OCs, \citet{Joshi_2016} derived the colour-excess, distance, age and total mass of Trumpler 2 as $E(B-V)=0.321$ mag, $d=670$ pc, $\log{t}$ = 7.925 yr, and $M/M_{\odot}=72.63\pm 0.19$, respectively. With the beginning of the {\it Gaia} era \citep{Gaia_DR1}, the quality and accuracy of photometric, astrometric, and spectroscopic data have considerably taken a turn for the better. \citet{Cantat-Gaudin_2018} investigated 1229 open clusters using the second data release of {\it Gaia} \citep[{\it Gaia} DR2;][]{Gaia_DR2} and listed the mean astrometric parameters and membership probabilities of open clusters to characterize them. \citet{Cantat-Gaudin_2018} identified 179 likely member stars ($P\geq 0.5$) in the direction of Trumpler 2, resulting is the proper-motion components being estimated as ($\mu_{\alpha}\cos\delta$, $\mu_{\delta}$) = (1.583 $\pm$ 0.018, -5.352 $\pm$ 0.015) mas yr$^{-1}$ and distance $d=685^{+50}_{-14}$ pc. \citet{Soubiran_2018} performed kinematic analyses for 861 open clusters using {\it Gaia} DR2 data, estimating the radial velocity $V_{\gamma}$ = -4.12 $\pm$ 0.09 km s$^{-1}$, and space velocity components ($U, V, W$) = (-4.12 $\pm$ 0.08, -11.98 $\pm$ 0.08, -13.60 $\pm$ 0.06) km s$^{-1}$ for Trumpler 2. Calculating the age, distance, distance modulus, and extinction values of 269 open clusters with the data obtained from the {\it Gaia} DR2 catalogue using the Bayesian statistical method, \citet{Bossini_2019} determined the age of Trumpler 2 cluster  as $91^{+14}_{-7}$ Myr. \citet{Liu_2019}  used 86 member stars to calculate the mean proper-motion components, trigonometric parallaxes, and age. Accordingly, the values of the parameters as given as ($\langle \mu_{\alpha}\cos\delta$, $\mu_{\delta}\rangle$) = (1.574 $\pm$ 0.273, -5.323 $\pm$ 0.242) mas yr$^{-1}$, $\langle \varpi \rangle= 1.441 \pm 0.048$ mas, and $t$ = 81 $\pm$ 5 Myr. \citet{Cantat-Gaudin_2020} obtained the astrometric and astrophysical parameters belonging to 2017 open clusters  based on {\it Gaia} DR2 data. According to this study, the mean proper-motion components and parallaxes are calculated from 154 member stars as ($\langle \mu_{\alpha}\cos\delta$, $\mu_{\delta}\rangle$) = (1.583 $\pm$ 0.192, -5.352$\pm$ 0.167) mas yr$^{-1}$ and $\langle \varpi \rangle= 1.431 \pm 0.058$ mas, respectively. The recent study on the Trumpler 2 was carried out by \citet{Carrera_2022}. They performed high-resolution spectroscopic analyses and obtained metallicities for 41 stars with high membership probabilities in 20 open clusters including Trumpler 2. They determined metallicity value of Trumpler 2 as -0.262 $\pm$ 0.106 dex. Furthermore, analysing {\it Gaia} EDR3 astrometric data, \citet{Carrera_2022} obtained birth radius as  8.08 $\pm$ 0.05 kpc as well as the kinematic and dynamic orbital parameters of the cluster. The literature results are listed in Table \ref{tab:literature} for ease of comparison.

\begin{table*}
{\small

\setlength{\tabcolsep}{5pt}
\renewcommand{\arraystretch}{1.5}
  \centering
  \caption{Astrophysical parameters for Trumpler 2 compiled from the literature: Colour excesses ($E(B-V$)), distance moduli ($\mu_{\rm V}$), distances ($d$), iron abundances ([Fe/H]), age ($t$), proper-motion components ($\langle\mu_{\alpha}\cos\delta\rangle$, $\langle\mu_{\delta}\rangle$), radial velocity ($V_{\gamma}$) and reference (Ref).}
  \begin{tabular}{ccccccccc}
    \hline
    \hline
$E(B-V)$ & $\mu_{\rm V}$ & $d$ & [Fe/H] & $t$ &  $\langle\mu_{\alpha}\cos\delta\rangle$ &  $\langle\mu_{\delta}\rangle$ & $V_{\gamma}$ & Ref \\
(mag) & (mag) & (pc)  & (dex) & (Myr) & (mas yr$^{-1}$) & (mas yr$^{-1}$) & (km s$^{-1})$ &      \\
    \hline
 0.40$\pm$0.06   & 10.50$\pm$0.35    & 725               & 0.00          & 89                     & $-$              & $-$              & $-$                          & (01) \\
 $-$             & $-$               &  $-$              & $-$           &  $-$                    & 1.0$\pm0.3$      & -4.6$\pm0.3$      & $-$                         & (02) \\
 $-$             & $-$               & $-$               & $-$           & $-$                     & 0.11$\pm1.48$    & -3.49$\pm2.23$    & $-$                         & (03)\\
 0.321           & $-$            & 670               & $-$           & 84                      & $-$              & $-$               & $-$                         & (04) \\  
 $-$             & $-$               & $685_{-44}^{+50}$ & $-$           & $-$                     & 1.583$\pm$0.018  &  -5.352	$\pm$0.015    & $-$                          & (05) \\
 $-$             & $-$               & $685_{-44}^{+50}$ & $-$           & $-$                     & 1.583$\pm$0.018  & -5.352	$\pm$0.015 & -4.12$\pm$0.09              & (06) \\
$0.308_{-0.006}^{+0.007}$             & 10.289$\pm$0.021  & 736$\pm$7            & 0.00          & $91_{-7}^{+14}$         & $-$              & $-$               & $-$                         & (07) \\
$-$              & $-$               & 694$\pm$23       & $-$           & 81$\pm$5                      & 1.574$\pm$0.273  & -5.323$\pm$0.242  & $-$                         & (08) \\
0.277            & 10.12            & 710         & $-$           & 112                     & 1.583$\pm$0.192  & -5.352	$\pm$0.167 & $-$                      &(09) \\ 
0.307           &  $-$           & 685              & -0.088$\pm${0.07}        & 92          &1.583$\pm$0.192    &-5.352$\pm$0.167     &-6.045$\pm$12.32         & (10) \\
 
 0.355$\pm$0.041 & 10.25$\pm$0.029   & 677$\pm$9    & 0.140$\pm$0.149 & 238$\pm$132          & 1.581$\pm$0.236 & -5.343$\pm$0.193                  & -3.990$\pm$0.746        & (11) \\
 
 0.316   & $-$  &  710 & -0.262$\pm$0.106 & 110          & {1.58$\pm$0.19} & -5.35$\pm$0.17 & -4.16$\pm$0.02 & (12) \\

  \hline
    \end{tabular}%
    \\
(01) \citet{Frolov_2006}, (02) \citet{Zejda_2012}, (03) \citet{Dias_2014}, (04) \citet{Joshi_2016}, (05) \citet{Cantat-Gaudin_2018}, (06) \citet{Soubiran_2018}, (07) \citet{Bossini_2019}, (08) \citet{Liu_2019}, (9) \citet{Cantat-Gaudin_2020}, (10) \citet{Zhong_2020}, (11) \citet{Dias_2021}, (12) \citet{Carrera_2022} 
  \label{tab:literature}%
}
  
\end{table*}%

\section{Data}
\subsection{Astrometric and Photometric Data}

\begin{figure*}
\centering
\includegraphics[scale=0.90, angle=0]{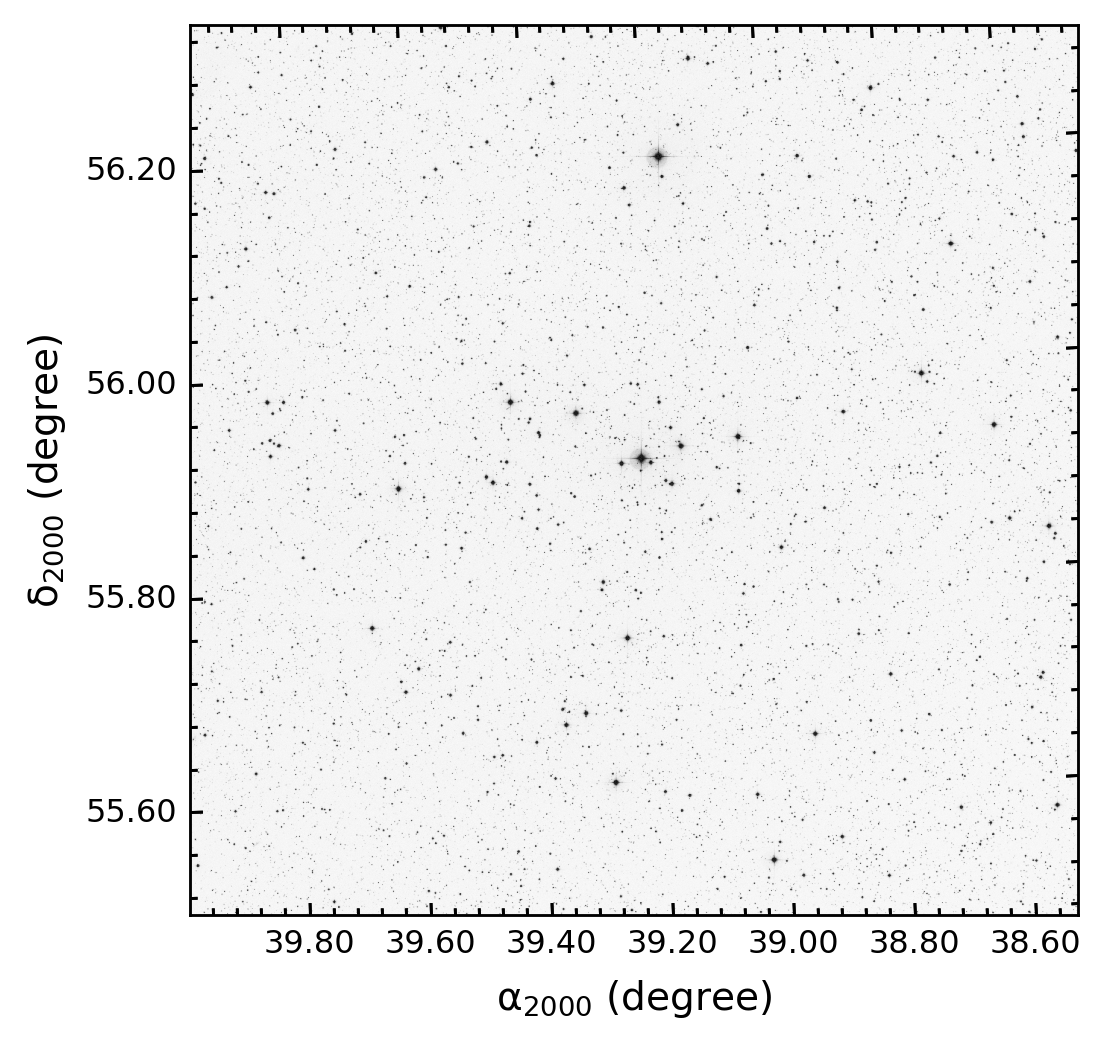}
\caption{Identification chart of Trumpler 2. The field of view of the chart is $50' \times 50'$. North and East correspond with the up and left directions, respectively.} 
\label{fig:ID_charts}
\end {figure*}

By using the high-quality data of the third data release of {\it Gaia} \citep[{\it Gaia} DR3,][]{Gaia_DR3}, we constructed the astrometric and photometric catalog of the cluster. For this, we used the equatorial coordinates calculated by \citet{Cantat-Gaudin-Anders_2020} ($\langle\alpha, \delta\rangle) = (02^{\rm h} 36^{\rm m} 55^{\rm s}.7, \delta= +55^{\circ} 54^{\rm '} 18^{\rm''}$) and take into account all stars located throughout the field of Trumpler 2 within the 50 arcmin region around the centre. Therefore  it is found out that there are 121,569 stars with the stellar magnitudes between $7<G\leq23$ mag. The identification chart of stars ($50' \times 50'$) located in regions of Trumpler 2 is shown in Figure~\ref{fig:ID_charts}. 

\subsection{Photometric Completeness Limit and Photometric Errors}
\begin{figure*}
\centering
\includegraphics[scale=0.6, angle=0]{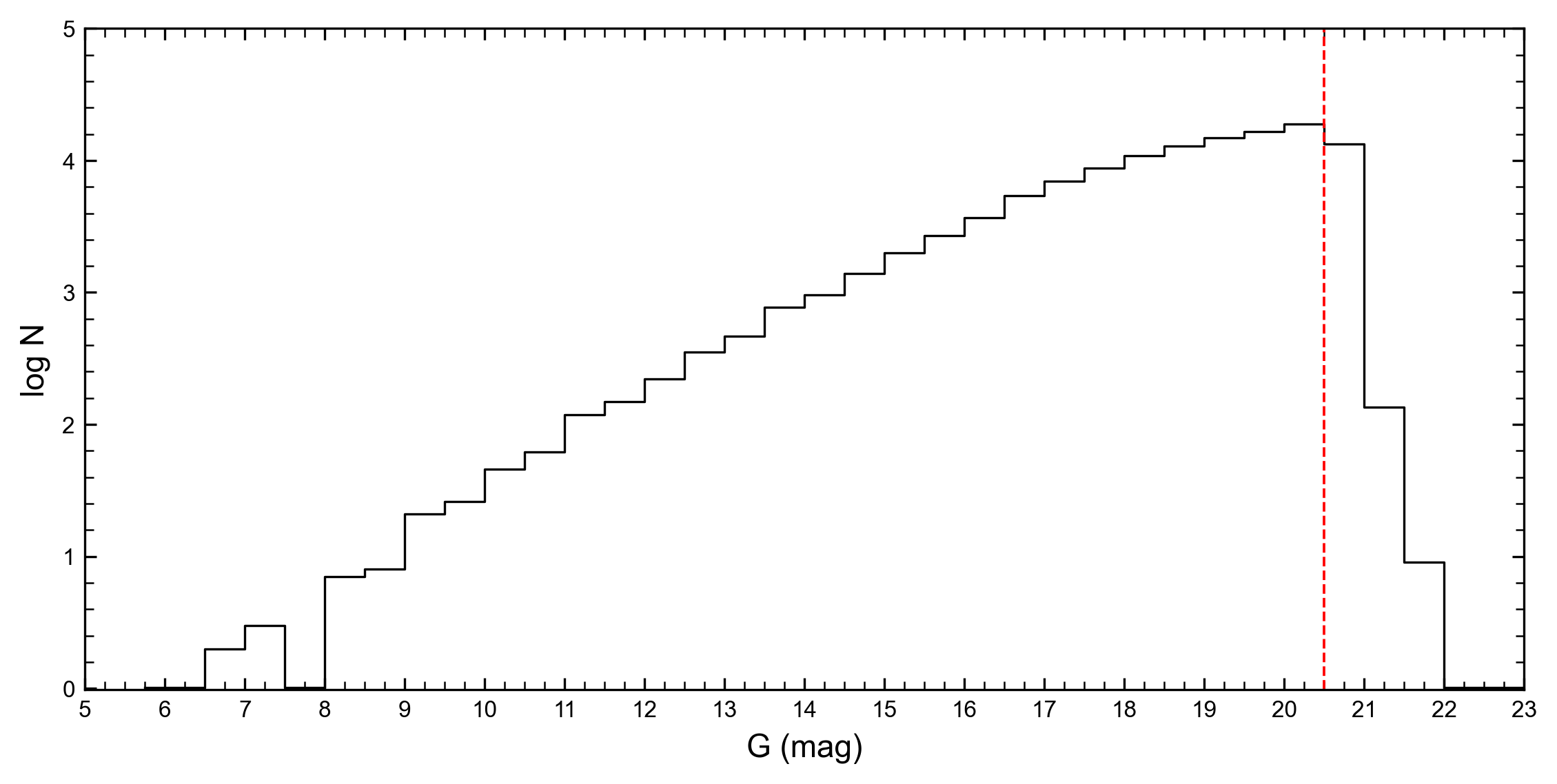}\\
\caption{Histogram of number of stars with respect to $G$ magnitude intervals for Trumpler 2. The red dashed-line indicates the faint limit magnitude of the cluster.}
\label{fig:histograms}
\end {figure*}

\begin{figure*}
\centering
\includegraphics[scale=1.02, angle=0]{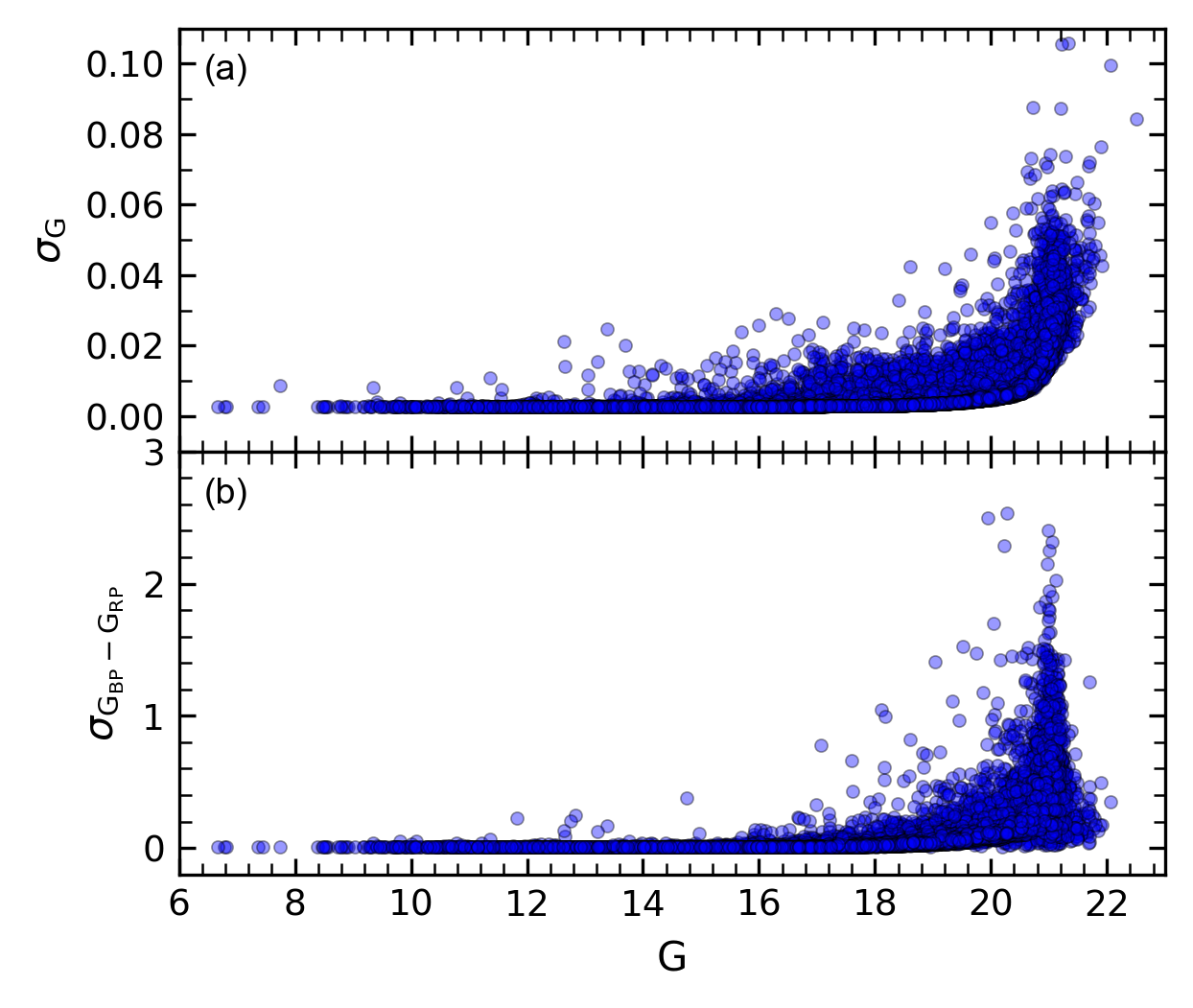}
\caption{~Mean photometric errors calculated for $G$ apparent magnitude (a) and $G_{\rm BP}-G_{\rm RP}$ colour index for Trumpler 2.} 
\label{fig:photometric_errors}
\end {figure*}

 To calculate reliable structural and astrophysical parameters of open clusters, it is important to obtain the photometric completeness limit of the data. To determinate the accurate parameter estimates for Trumpler 2, the photometric completeness limit  of the stars is established by counting the stars corresponding to the $G$ magnitudes. In Figure~\ref{fig:histograms}, the histogram shows the completeness limit as $G=20.50$ mag since the count of stars increases with increasing magnitude until this value and then decreases after it. In the following analyses, stars fainter than this value were excluded statistically and were not taken into account. The uncertainties in the {\it Gaia} DR3 data were accepted as interval errors and calculated mean $G$ magnitudes with $G_{\rm BP}-G_{\rm RP}$ color indices of stars in the cluster region as a function of $G$ magnitude intervals. Considering $G=20.50$ mag , the mean internal $G$ magnitude and $G_{\rm BP}-G_{\rm RP}$ colour index were calculated as 0.007 and 0.142 mag at this limit, respectively. Mean errors of photometric data are listed in Table \ref{tab:photometric_errors} as a function of $G$ apparent magnitude. Also, Figure \ref{fig:photometric_errors} presents the mean internal errors of $G$ magnitudes and $G_{\rm BP}-G_{\rm RP}$ colour indices.
 
\begin{table}
\setlength{\tabcolsep}{10pt}
\renewcommand{\arraystretch}{1.3}
  \centering
  \caption{ Mean internal photometric errors per magnitude bin in $G$ apparent magnitude and $G_{\rm BP}-G_{\rm RP}$ colour index for Trumpler 2.}
    \begin{tabular}{cccc}
      \hline
  $G$ (mag) & $N$ & $\sigma_{\rm G}$ (mag) & $\sigma_{G_{\rm BP}-G_{\rm RP}}$ (mag) \\
                                                             
  \hline
    (06, 14]	&  1839  & 0.003  & 0.006 \\
    (14, 15]	&  2056	&  0.003  &	0.006 \\
    (15, 16]	&  3930	&  0.003  &	0.006 \\
    (16, 17]	&  7566	&  0.003  &	0.009 \\
    (17, 18]	& 14083	&  0.003  &	0.017 \\
    (18, 19]	& 21699	&  0.003  &	0.038 \\
    (19, 20]	& 29495	&  0.005  &	0.081 \\
    (20, 21]	& 37454	&  0.010  &	0.208 \\
    (21, 23]	& 3447	&  0.027  &	0.430 \\
   \hline
    \end{tabular}%
  \label{tab:photometric_errors}%
\end{table}%

\section{Results}
\subsection{Structural Parameters of Trumpler 2}
\begin{figure}[]
\centering
\includegraphics[scale=0.20, angle=0]{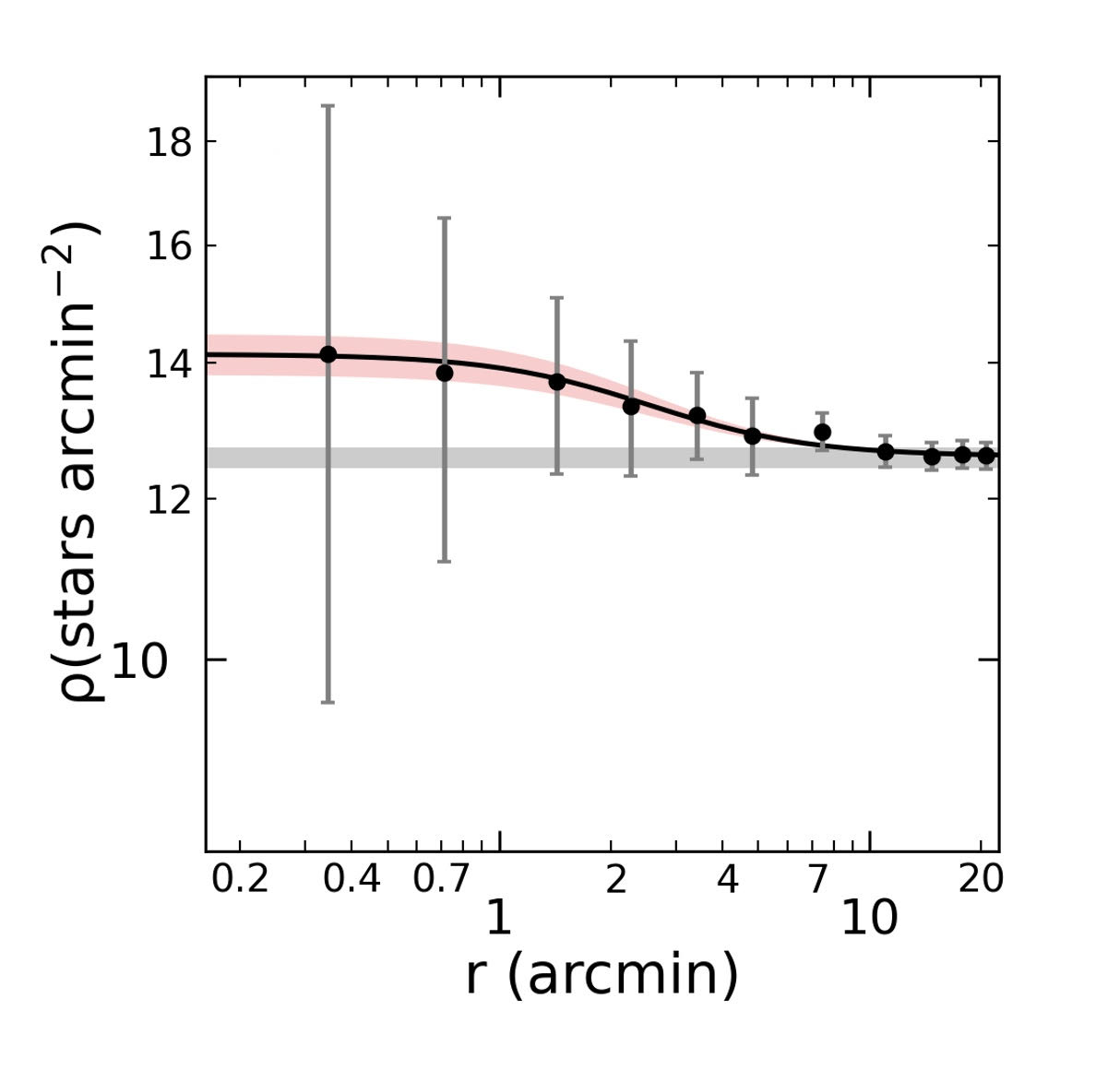}\\
\caption{The stellar density distribution of Trumpler 2. The fitted black curve shows the RDP profile of \citet{King_1962}, whereas the horizontal grey band represents the background stellar density. The $1\sigma$ King fit uncertainty is pictured by the red-shaded area.} 
\label{fig:king}
\end {figure}

In order to investigate the extent of Trumpler 2 and determine its
structural parameters, we performed Radial Density Profile (RDP) analysis. By using {\it Gaia} DR3 data within 50 arcmin, we divided the cluster area into many concentric rings from the central coordinates given by \citet{Cantat-Gaudin_2020}. For the calculation of the stellar density ($\rho(r)$), we used the stars within the $G \leq 20.5$ mag completeness limit and utilized the equation of $R_{i}=N_{i}/A_{i}$ for each $\it i^{\rm th}$ ring, where $N_{i}$ and $A_{i}$ are the number of stars falling into a given ring and the area of that ring, respectively. Uncertainties in the stellar density were calculated using the Poisson statistical measure $1/\sqrt N$, where $N$ is the number of stars. Then, to plot the radial density profile we used the stellar density and fitted the RDP with the empirical \citet{King_1962} model defined by the following formula: 

\begin{equation}
\rho(r)=f_{\rm bg}+\frac{f_0}{1+(r/r_{\rm c})^2}
\end{equation}
In the equation, $r$ denotes the radius of the cluster. The terms of  $f_{\rm bg}$, $f_0$, $r_{\rm c}$ represent the background stellar density, the central stellar density and the core radius of Trumpler 2, respectively. $\chi^{2}$ minimisation technique was used in the RDP fitting process to estimate the $f_{\rm bg}$, $f_0$ and $r_{\rm c}$ parameters which are listed in Table \ref{tab:rdp}. The best-fit solution to the radial density profile is illustrated with a solid black line in Figure \ref{fig:king}. The correlation coefficient of the King model \citep{King_1962} with the best-fit model parameters to the data is $R^2=0.971$, which implies that the estimated structural parameters are well estimated.

Taking into account the visualisation of RDP, we determine the limiting radii ($r_{\rm lim}^{\rm obs}$) of cluster as $12$. The structural parameters were derived as $f_{\rm bg}=12.59\pm0.02$, $f_0= 1.55\pm0.33$ stars arcmin$^{-2}$ and $r_{\rm c}=2.48\pm1.09$ arcmin (see also Table \ref{tab:rdp}). To validate the precision of the observed limiting radii ($r_{\rm lim}^{\rm obs}$), 
equation given by \citet{Bukowiecki_2011} is used as follow:

\begin{equation}
r_{\rm lim}=r_{\rm c}\sqrt{\frac{f_0}{3\sigma_{\rm bg}}-1}
\end{equation}

From the equation, we calculated the theoretical limiting radius as 11.7 arcmin. It is obvious that theoretical and observed limiting radii values are in a good agreement with each other. 

\newcommand{\fscale}{(stars arcmin$^{-2}$)}
\begin{table}
\small
\setlength{\tabcolsep}{5.pt}
\renewcommand{\arraystretch}{1.1}
  \centering
  \caption{The structural parameters of the Trumpler 2 according to the \citet{King_1962} model analyses: $(\alpha, \delta)$, $f_0$, $f_{\rm bg}$, $r_{\rm c}$, $r_{\rm lim}^{\rm obs}$, $r_{\rm lim}^{\rm cal}$ and $R^2$ are equatorial coordinates, central stellar density, the background stellar densities, the core radius, observational limiting radius, calculated limiting radius and correlation coefficient, respectively.}
    \begin{tabular}{lcccccccc}
\hline
  Cluster & $\alpha_{\rm J2000}$ & $\delta_{\rm J2000}$  & $f_0$   & $f_{\rm bg}$ & $r_{\rm c}$ & $r_{\rm lim}^{\rm obs}$ & $r_{\rm lim}^{\rm cal}$& $R^2$\\ 
          & (hh:mm:ss.s)       &   (dd:mm:ss)        & \fscale &  \fscale     &  (arcmin)   &  (arcmin) & (arcmin)   & \\
\hline
Trumpler ~~ 2 & 02:36:55.7 & +55:54:18 & 1.55$\pm$0.33 & 12.59$\pm$0.02 & 2.48$\pm$1.09 & ~~12  & ~~11.7 & 0.971\\

\hline
    \end{tabular}%
    \label{tab:rdp}%
\end{table}%

\subsection{Membership Probabilities of Stars}

As open clusters are located within the Galactic plane, they highly contaminated by foreground and background stars. This situation makes it necessary to separate cluster members from field stars for determining reliable astrophysical parameters of the studied cluster. As the member stars of an OC form under similar physical conditions, they have similar vectorial movements in space. These properties make proper-motion components functional information to distinguish cluster members from the field stars \citep{Bisht_2020}. The high accuracy of {\it Gaia} DR3 data \citep{Gaia_DR3} leads to reliable results for membership analyses. To determine the membership probabilities ($P$) of stars in the direction of Trumpler 2, we used the membership determination method known as Photometric Membership Assignment in Stellar Clusters \citep[UPMASK,][]{Krone-Martins_2014} on the Gaia DR3 astrometric data. Many researchers have previously utilized this method \citep{Cantat-Gaudin_2018, Cantat-Gaudin_2020, Castro-Ginard_2018, Castro-Ginard_2019, Castro-Ginard_2020, Banks_2020, Akbulut_2021, Wang_2022, Yontan_2023}. {\sc upmask} is a clustering method which is defined as $k$-means. The method allows the detection of similar groups of stars according to their proper-motion components and trigonometric parallax, providing statistical probability values for membership of these groups. We used each detected stars' astrometric measurements ($\alpha$, $\delta$, $\mu_{\alpha}\cos \delta$, $\mu_{\delta}$, $\varpi$) as well as their uncertainties by running 100 iterations of {\sc upmask} to set their membership probability values. Hence, we identified 92 stars for Trumpler 2 as the most probable member stars with membership probabilities $P\geq 0.5$, brighter than the photometric completeness limits and located within the limited radius ($r_{\rm lim}^{\rm obs}$) described above.  

\begin{figure}
\centering
\includegraphics[scale=0.50, angle=0]{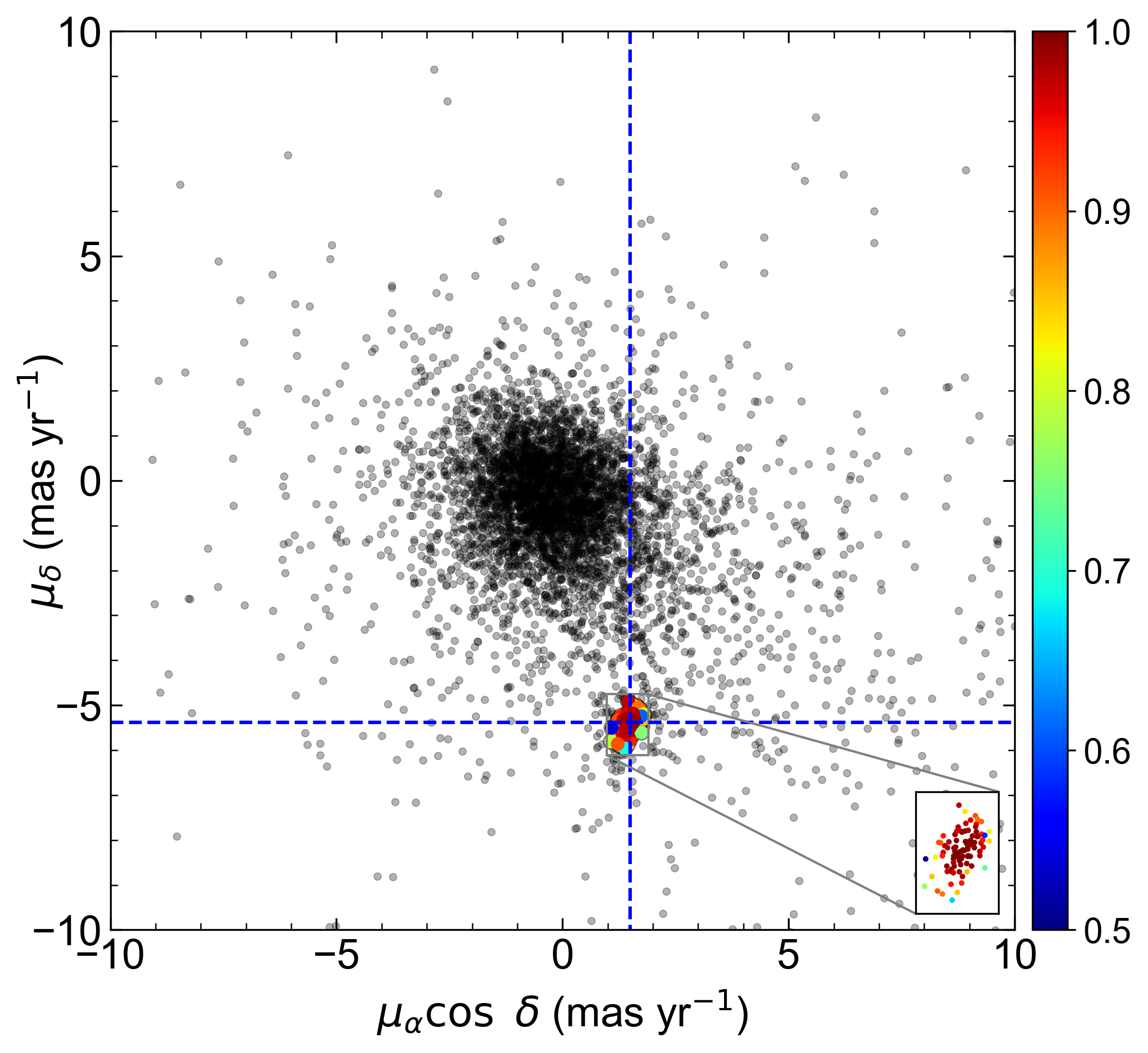}\\
\caption{VPD of Trumpler 2 based on {\it Gaia} DR3 astrometry. The membership probabilities of the stars are identified with the colour scale shown on the right for the cluster. The zoomed in panel inset represents the region of condensation for the cluster in the VPD. The mean proper-motion value of Trumpler 2 is shown by the intersection of the dashed blue lines.
\label{fig:VPD_all}} 
 \end {figure}

\begin{figure}
\centering
\includegraphics[scale=.50, angle=0]{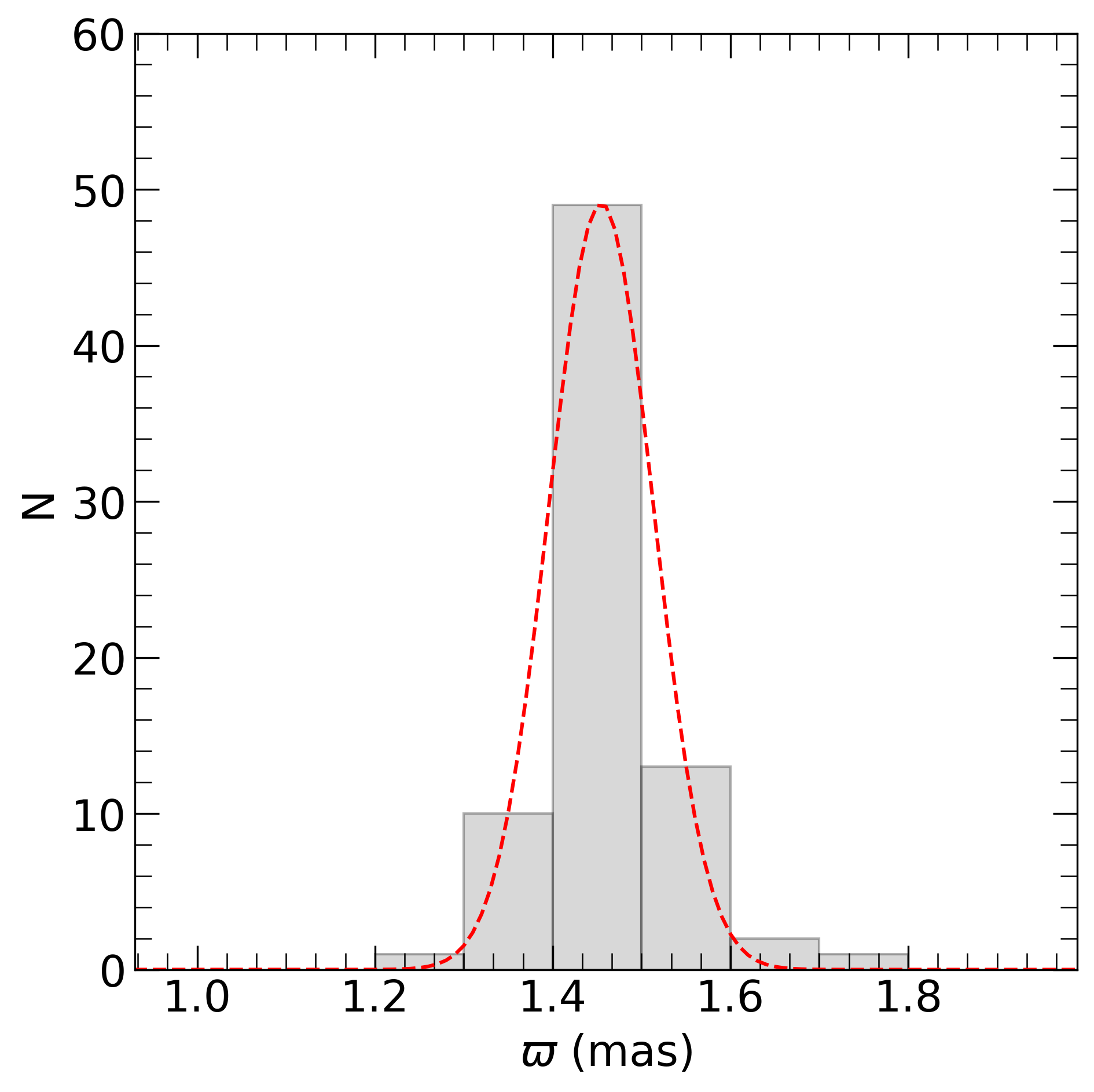}\\
\caption{Histogram of mean trigonometric parallax estimation of Trumpler 2 from the most probable members ($P\geq 0.50$). Red dashed line shows the fitted Gaussian function.
\label{fig:plx_hist}}
\end {figure}

We constructed the Vector-Point Diagram (VPD), which consists of proper-motion components of the stars, to investigate the most probable cluster member stars, as shown in Figure ~\ref{fig:VPD_all}. It can be seen from the figure that Trumpler 2 is clearly distinguished from the foreground and background stars. In Figure ~\ref{fig:VPD_all}, the intersection of the blue dashed lines represents the mean proper-motion values obtained from the most probable member stars ($P\geq0.5$). The mean proper-motion components of Trumpler 2 are found as ($\mu_{\alpha}\cos \delta$, $\mu_{\delta}$)=($1.494 \pm 0.004$, $-5.386 \pm 0.005$) mas yr$^{-1}$. Also, using these member stars we plotted the trigonometric parallax histogram and calculated the mean value by fitting the Gaussian functions to the distributions as shown in Figure~\ref{fig:plx_hist}. Thus, we estimated  the mean trigonometric parallax value as $\varpi$=1.455$\pm 0.059$ mas for the cluster. We transformed the mean trigonometric parallax to the trigonometric parallax distance ($d_{\varpi}$) by using the linear equation of $d({\rm pc})=1000/\varpi$ (mas) and estimated the value as $d_{\varpi}=687\pm 27$ pc. This result is in a good agreement with the results determined in the {\it Gaia} era by different researchers (for a detailed comparison see Table~\ref{tab:literature}).

\subsection{Astrophysical Parameters of Trumpler 2}
Colour-magnitude diagrams (CMDs) are essential tools to obtain basic astrophysical parameters such as age, distance modulus, reddening and metallicity of open clusters. These diagrams allow identification of a cluster's main sequence, turn-off point, and giant members as well as the determination of reliable cluster parameters. In the current study, considering the morphology of the most probable ($P\geq 0.5$) cluster members on  the {\it Gaia} based CMD, we estimated age, distance moduli and reddening simultaneously, whereas the metallicity [Fe/H] was adopted with the value of [Fe/H]=$-0.262\pm 0.106$ dex as given by \citet{Carrera_2022}. To derive the age, distance moduli and reddening of Trumpler 2, we applied {\sc PARSEC} stellar isochrones \citep{Bressan_2012} to the observed CMD and obtained the best fit. We used the {\sc PARSEC}\footnote{http://stev.oapd.inaf.it/cgi-bin/cmd} isochrones corresponding to {\it Gaia} Early Data Release 3 (EDR3) photometric passbands \citep{Riello_2021}. 

We transformed the adopted metallicity [Fe/H]=$-0.262\pm 0.106$ dex) to the mass fraction $z$ for the selection of isochrones as well as for determination of astrophysical parameters. For the transformation, we applied the analytic expressions of Bovy\footnote{https://github.com/jobovy/isodist/blob/master/isodist/Isochrone.py} suitable for {\sc parsec} isochrones \citep{Bressan_2012}. The expressions are given as follow:

\begin{equation}
z_{\rm x}={10^{{\rm [Fe/H]}+\log \left(\frac{z_{\odot}}{1-0.248-2.78\times z_{\odot}}\right)}}
\end{equation}      
and
\begin{equation}
z=\frac{(z_{\rm x}-0.2485\times z_{\rm x})}{(2.78\times z_{\rm x}+1)}.
\end{equation} 
where $z_{\rm x}$ and $z_{\odot}$ indicate intermediate values where solar metallicity $z_{\odot}$ was accepted as 0.0152 \citep{Bressan_2012}. We derived $z=0.0088$ for Trumpler 2.

To determine the age, distance moduli and reddening of the cluster, we constructed  a $G\times (G_{\rm BP}-G_{\rm RP})$ CMD and fitted selected isochrones by visual inspection, according to the most probable ($P\geq 0.5$) main sequence, turn-off, and giant member stars. Before the selection of isochrones, we scaled them to the mass fraction $z$, as determined above. This procedure indicated that the {\it Gaia}-based colour excess for Trumpler 2 is $E(G_{\rm BP}-G_{\rm RP})$= $0.452\pm 0.019$ mag. To compare with literature studies more accurately, we transformed this value to the $U\!BV$-based colour excess $E(B-V)$ by employing the equation $E(G_{\rm BP}-G_{\rm RP})= 1.41\times E(B-V)$ as given by \citet{Sun_2021}. Hence, we find the value as $E(B-V)$=$0.320\pm 0.013$ mag which is compatible with the results given by \citet{Joshi_2016}, \citet{Bossini_2019}, \citet{Zhong_2020} and \citet{Carrera_2022} within the errors (see Table~\ref{tab:literature}). 

\begin{figure}
\centering
\includegraphics[scale=0.59, angle=0]{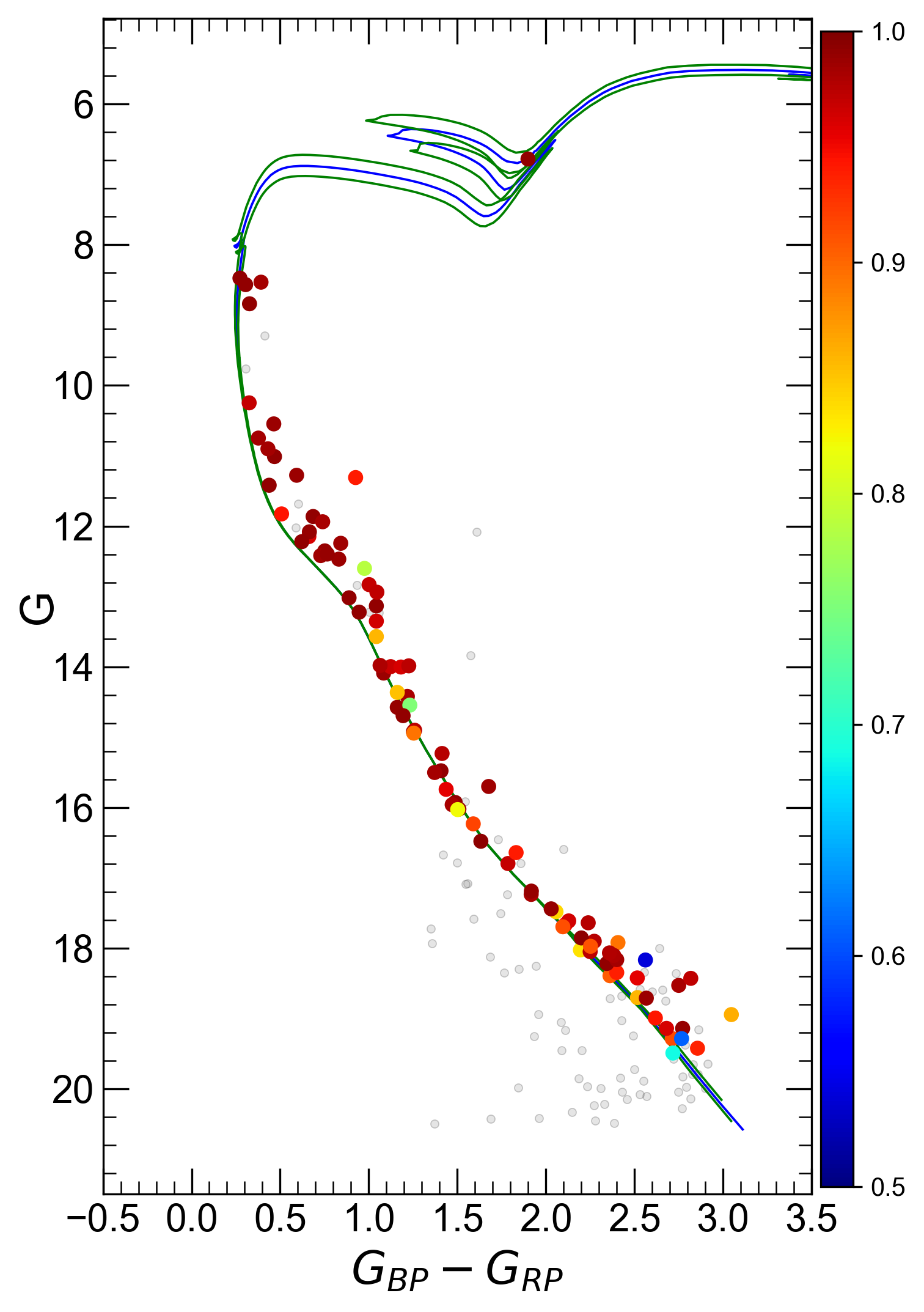}
\caption{Colour-magnitude diagram for the studied cluster Trumpler 2. Different colour scales show the membership probabilities of the most probable cluster members. The scales of memberships are represented on colourbars to the right. Grey coloured dots identify the stars with probabilities $P<0.5$. The best fitting {\sc parsec} isochrones and their errors are shown as the blue and  green lines, respectively. Superimposed isochrone ages match to 110 Myr for Trumpler 2
\label{fig:figure_age} }
\end {figure}

The {\sc PARSEC} isochrones with different ages of $\log({\rm age}) = 8.00, 8.04$ and 8.08 with $z=0.0088$ were superimposed to the observed $G\times (G_{\rm BP}-G_{\rm RP})$ CMD as shown in Figure ~\ref{fig:figure_age}. The overall fit is acceptable for $\log({\rm age})$ = 8.04 to the most probable members ($P\geq 0.5$), corresponding to the age of Trumpler 2 being $t=110\pm 10$ Myr. The derived  age value is compatible with the  values given by \citet{Bossini_2019}, \citet{Zhong_2020} and \citet{Cantat-Gaudin_2020} within the formally stated errors (Table~\ref{tab:literature}). The estimated distance modulus $\mu=10.027\pm 0.149$ mag  corresponds to an isochrone distance from Sun as $d_{\rm iso}=686\pm 49$ pc. The errors in distance modulus and isochrone distance were calculated from the relations presented in \citet{Carraro_2017}. These relations take into account the photometric magnitude and colour excess with their uncertainties. The estimated isochrone distance is in a good agreement with the values given by different researchers (see Table~\ref{tab:literature}) as well as with the value of trigonometric parallax distance $d_{\varpi}=687\pm 27$ pc calculated in this study.

Moreover, we derived Galactocentric coordinates $(X, Y, Z)_{\odot}$  of Trumpler 2, where $X$, $Y$, and $Z$ are directions toward the Galactic centre in the Galactic disc, the Galactic rotation and the Galactic north pole, respectively. We used isochrone distance, Galactic longitude and latitute of the cluster and calculated these values as $(X, Y, Z)_{\odot}=$($-504$, 464, -48) pc. These findings are in fair agreement with the results of \citet{Cantat-Gaudin_2020} (see also Table~\ref{tab:Final_table}).

\section{Space Velocities and Galactic Orbital Parameters}

Kinematic and dynamical orbit analyses of  open clusters are important to understand the birth radii and Galactic populations of these objects \citep{Yontan_2023}. In this study we performed orbital integration analyses and obtained the Galactic orbital parameters for Trumpler 2. To do this, we used {\sc MWPotential2014} in  the Galactic dynamics library {\sc galpy}\footnote{See also https://galpy.readthedocs.io/en/v1.5.0/} package of \citet{Bovy_2015}  which is written in Python\footnote{See also https://galpy.readthedocs.io/en/v1.5.0/}. The {\sc MWPotential2014} model represents the sum of three-component potentials: the spherical bulge as described in \citet{Bovy_2015}, the Galactic disc as defined by \citet{Miyamoto_1975} and the massive, spherical dark-matter halo as defined by  \citet{Navarro_1996}. The Galactocentric distance and orbital velocity of Sun were taken as $R_{\rm gc}=8$ kpc and $V_{\rm rot}=220$ km s$^{-1}$, respectively \citep{Bovy_2015, Bovy_2012}.  The distance from the Galactic plane of the Sun was accepted as $25 \pm 5$ pc \citep{Juric_2008}.

For the complete integration of the orbit, radial velocity information is one of the necessary parameter to be known. We used the available {\it Gaia} DR3 radial velocities of the stars to determine mean radial velocity and its uncertainty for Trumpler 2. To do this, we took into account the most probable member stars with membership probabilities $P\geq 0.9$, and the number of these stars are 33. The calculation was based on the weighted average of stars' radial velocities \citep[for equations see ][]{Soubiran_2018}. Thus, the mean radial velocity was obtained as $V_{\gamma}= -4.48 \pm 1.11$ km s$^{-1}$. The literature mean radial velocity findings for Trumpler 2 (see also Table~\ref{tab:literature}) are $-4.12 \pm 0.09$ \citep{Soubiran_2018}, $-6.045 \pm 12.32$  \citep{Zhong_2020}, $-3.990 \pm 0.746$ km s$^{-1}$ \citep{Dias_2021} and $-4.16 \pm 0.02$ km s$^{-1}$ \citep{Carrera_2022}. The mean radial velocity obtained in this study for Trumpler 2 is within ~1 km s$^{-1}$ of the values are presented in these literature studies.                       

To carry out the orbit integration, we provide the following parameters as input: the central equatorial coordinates ($\alpha=02^{\rm h} 36^{\rm m} 55^{\rm s}.7$, $\delta= +55^{\rm o} 54^{\rm '} 18^{\rm''}$) taken from \citet{Cantat-Gaudin_2020}, the mean proper-motion components ($\mu_{\alpha}\cos\delta = 1.494\pm0.004$, $\mu_{\delta}= -5.386\pm0.005$ mas yr$^{-1}$) estimated in Section 3.2, the isochrone distance ($d_{\rm iso}=686\pm 49$ pc) from Section 3.3, and the radial velocity ($V_{\gamma}=-4.48\pm 1.11$ km s$^{-1}$) calculated in the study (see also Table~\ref{tab:Final_table}). 

\begin{figure*}
\centering
\includegraphics[scale=0.5, angle=0]{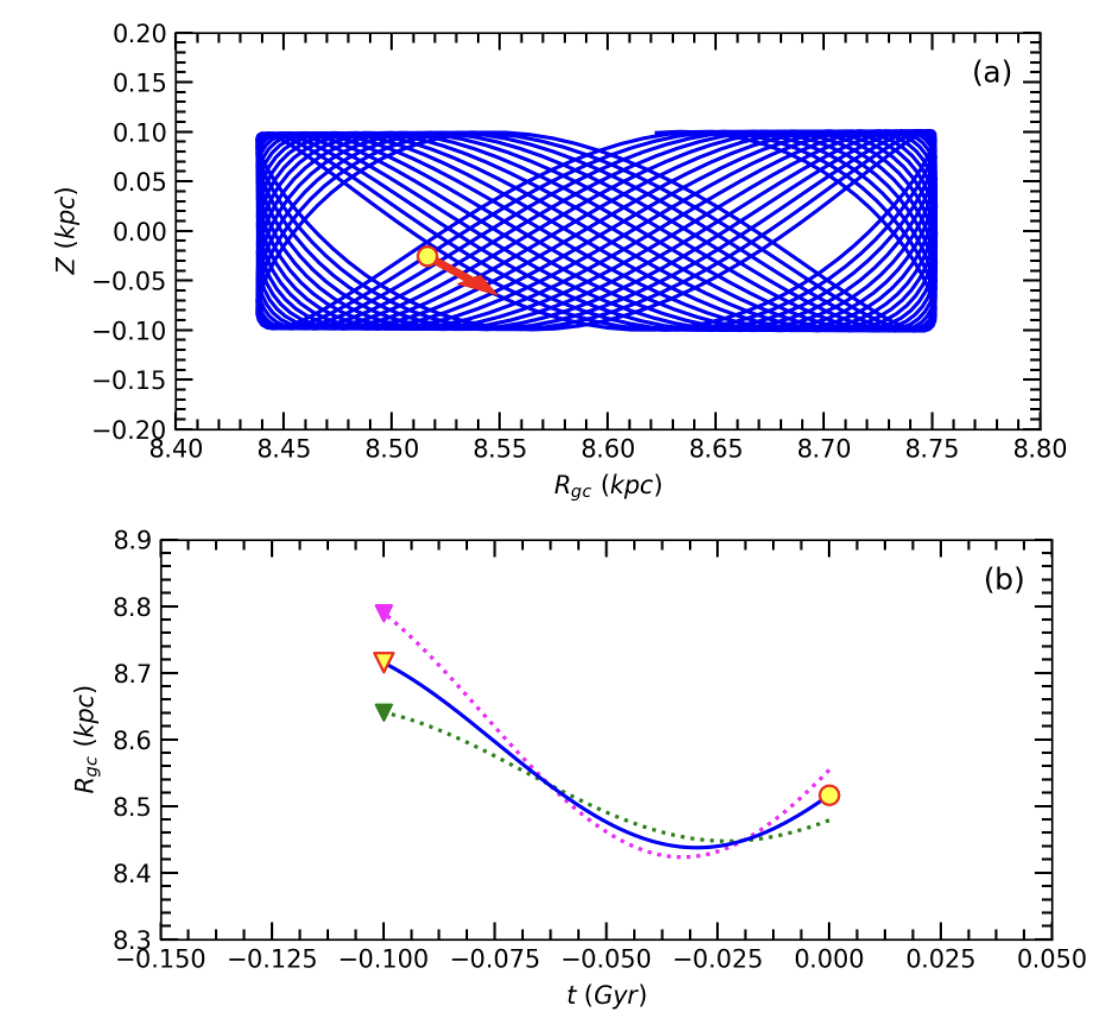}
\small
\caption{The Galactic orbits and birth radii of Trumpler 2 in the $Z \times R_{\rm gc}$ (a) and $R_{\rm gc} \times t$ (b) planes. The filled yellow circle and triangles show the present-day and birth positions, respectively. Red arrow is the motion vector of Trumpler 2. The green and pink dotted lines show the orbit when errors in input parameters are considered, while the green and pink filled triangles represent the birth locations of the open cluster based on the lower and upper error estimates.
\label{fig:galactic_orbits}}
\end {figure*}

We integrated the cluster orbit forward with an integration step of 1 Myr up to 2.5 Gyr in order to derive the cluster's possible current location. Figure~\ref{fig:galactic_orbits}a shows the  \textquotesingle side view\textquotesingle of the cluster on $Z \times R_{\rm gc}$ plane, which  indicates distance from the Galactic plane and the Galactic center. To estimates the possible birth radius of the Trumpler 2, we performed orbit analyses in the past epoch across a time equal to its age (110 Myr). The integration procedure was not carried out more than the cluster's age as the potential based uncertainties in time, as well as the additional errors in distance, proper-motion components and radial velocity influence the reliability of the results \citep{Gaia_DR2, Sariya_2021}. Figure~\ref{fig:galactic_orbits}b indicates the distance of the cluster on the and $R_{\rm gc} \times t$ plane with time.  In the figure the effect of uncertainties in the input parameters on the orbit of Trumpler 2 is represented. The results show that Trumpler 2 has an uncertainty of about 0.08 kpc for its possible birth-radius. These results also indicated that the cluster was formed outside the solar vicinity with a birth radius of 8.71 kpc.

Orbit integration resulted in the following derivation for Trumpler 2: apogalactic ($R_{\rm a}=8752\pm 107$ pc) and perigalactic ($R_{\rm p}=8438\pm10$ pc) distances, eccentricity ($e=0.018\pm0.007$), maximum vertical distance from Galactic plane ($Z_{\rm max}=101\pm 17$ pc), space velocity components ($U,V,W$=$-3.71\pm 1.32$, $-12.08\pm 0.10$, $-13.81\pm 1.07$ km s$^{-1}$), and orbital period ($P_{\rm orb}=242\pm1$ Myr). Taking into account the space velocity component values $(U, V, W)_{\odot}=(8.83\pm 0.24, 14.19\pm 0.34, 6.57\pm0.21$) km s$^{-1}$ of \citet{Coskunoglu_2011}, we applied a Local Standard of Rest (LSR) correction to the $(U, V, W)$ components of Trumpler 2. Hence, we derived the LSR corrected space velocity components as $(U, V, W)_{\rm LSR}$ = ($5.12\pm1.34$, $2.11\pm0.35$, $-7.24\pm1.09$) km s$^{-1}$. Using these LSR results, we estimated the total space velocity as $S_{\rm LSR}=9.12\pm1.76$ km s$^{-1}$, which is compatible with the velocity value given for young thin-disc stars \citep{Leggett_1992}. Perigalactic and apogalactic distances show that orbit of Trumpler 2 is completely outside the solar circle (Figure~\ref{fig:galactic_orbits}a). The cluster rises  a maximum distance above the Galactic plane at $Z_{\rm max}=101\pm 17$ pc, indicates that Trumpler 2 belongs to the thin-disc component of the Milky Way \citep{Bilir_2006b, Bilir_2006c, Bilir_2008}.

\section{Luminosity and Mass Functions}
\begin{figure}
\centering
\includegraphics[scale=1.2, angle=0]{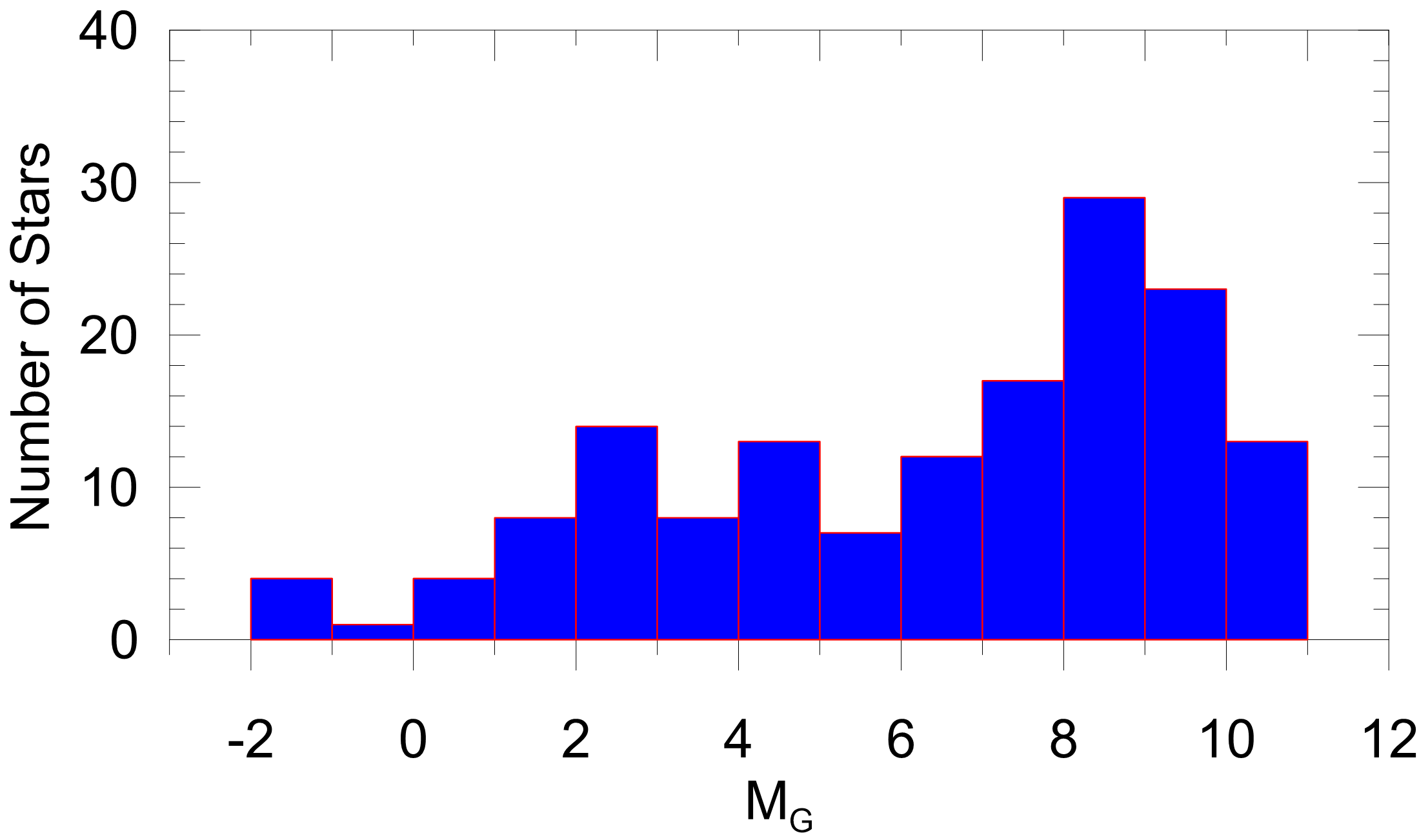}
\caption{\label{fig:luminosity_functions}
The luminosity function based on selected stars ($P>0$) for Trumpler 2.}
\end {figure}

The distribution of stellar brightness in an open cluster is known as its Luminosity Function (LF). The luminosity function and mass function (MF) are related by well known mass-luminosity relationships. The relationships between absolute magnitude and mass of the main-sequence stars can be estimated by a high-degree polynomial equation from the best fitted age isochrone to the cluster. We used main-sequence stars with probabilities $P>0$, located inside the limiting radius obtained in the study ($r_{\rm lim}^{\rm obs}=12$ arcmin) and  within the  magnitude range $10 \leq G \leq 19.5$ magnitudes to investigate the LF of Trumpler 2. With these selection criterias, LF analyses performed from 153 stars. We transformed apparent $G$ magnitudes of the selected stars into the absolute $M_{\rm G}$ magnitudes using the distance modulus equation $M_{\rm G} = G-5\times \log d +1.8626\times E(G_{\rm BP}-G_{\rm RP})$, where $G$, $d$ and $E(G_{\rm BP}-G_{\rm RP})$ are apparent magnitude, isochrone distance, and color excess that estimated for Trumpler 2 (see also Table~\ref{tab:Final_table}).  We plot as Figure~\ref{fig:luminosity_functions} the LF histogram, with a bin width of 1 magnitude. It can be seen from the figure that the absolute magnitude ranges of the used stars are lie within the $-2< M_{\rm G}< 11$ magnitude.  

The MF was estimated from the LF through the mass luminosity relation derived from the model isochrones \citep{Bressan_2012} based on the derived age and the previously adopted metallicity fraction ($z$) of the cluster. MF is described by the following expression:

\begin{eqnarray}
{\rm log(dN/dM)}=-(1+\Gamma)\times \log M + C.
\label{eq:mass_luminosity}
\end{eqnarray}
where $dN$ denotes the number of stars per unit mass $dM$, $M$ states the central mass, $C$ is the constant for the equation, and $\Gamma$ is the slope of the MF.  

\begin{figure}
\centering
\includegraphics[scale=1.2, angle=0]{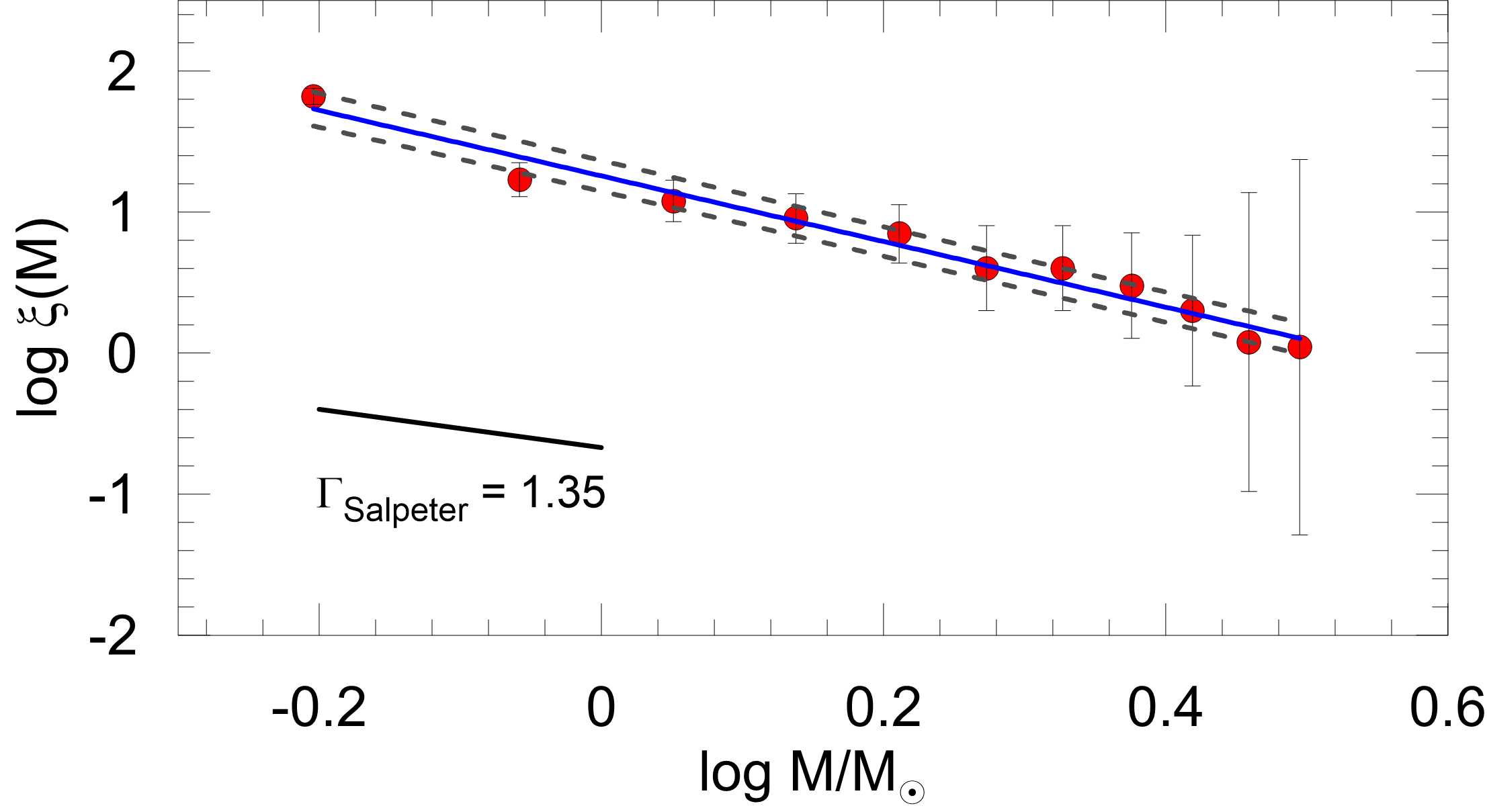}
\caption{\label{fig:mass_functions}
Mass function of Trumpler 2. Blue line indicates the calculated mass function of the cluster, whereas black dashed lines present the $\pm1\sigma$ standard deviations. Black solid line shows the slope of \citet{Salpeter_1955}.}
\end {figure}

We created a high-degree polynomial equation between the $G$-band absolute magnitudes and masses. Using this equation we transformed the observational absolute magnitudes $M_{\rm G}$ of the selected 153 stars ($P>0$) into masses. With this process we determined that the mass range of the 153 stars was inside $0.5\leq M/ M_{\odot}\leq 3.25$. Applying equation~\ref{eq:mass_luminosity}, we derived the MF slope value as $\Gamma=1.33 \pm 0.13$ for the cluster, as shown in Figure~\ref{fig:mass_functions}.

\begin{figure}
\centering
\includegraphics[scale=1.2, angle=0]{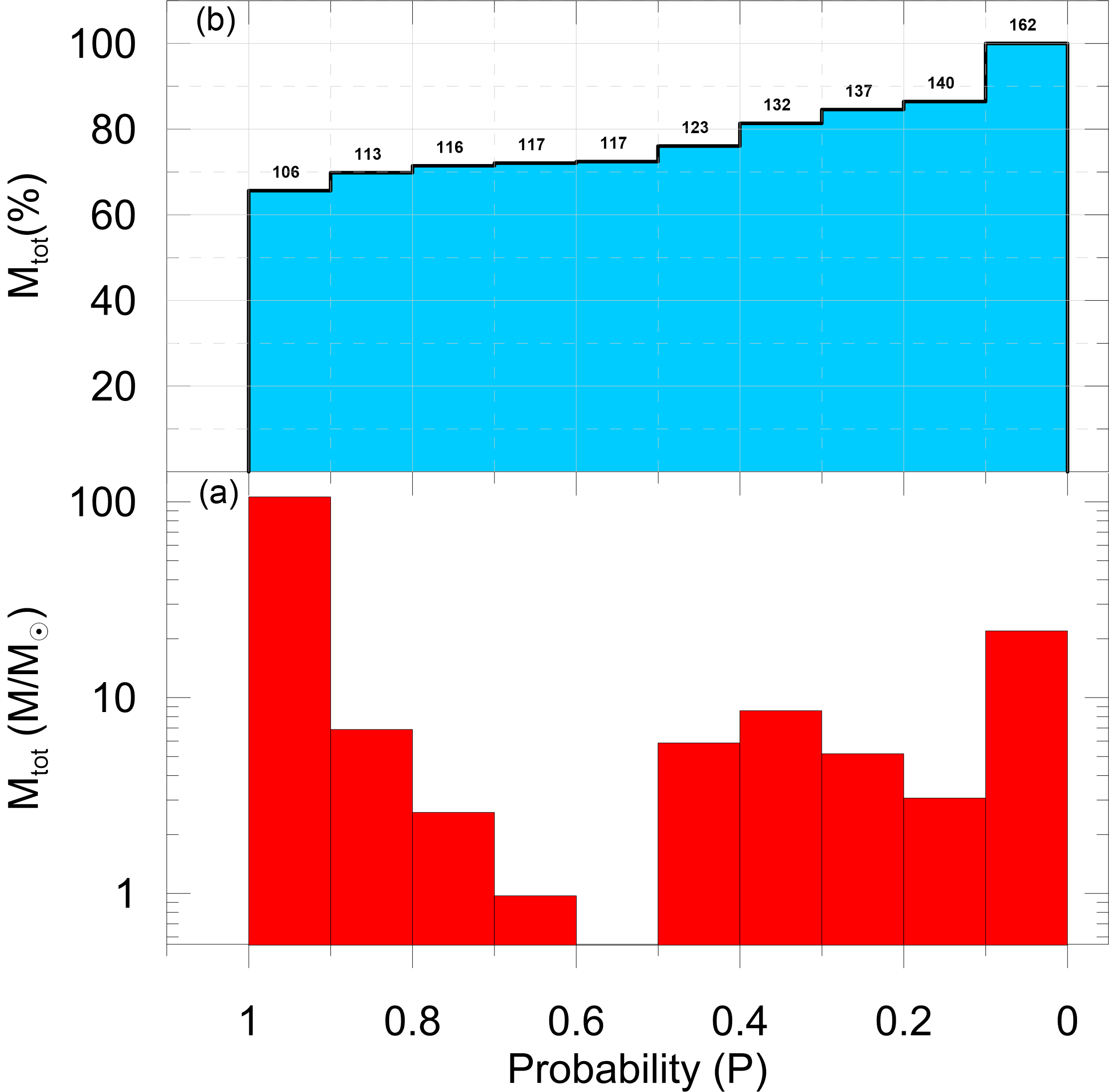}
\caption{\label{fig:total_mass}
Total mass of Trumpler 2 based on membership probabilities (a) and their cumulative masses (b). The numbers above the bins represent total mass of each bin in the histogram.}
\end {figure}

In the study, the total mass of the cluster is also investigated as a function of the membership probabilities of the stars, as shown in Figure~\ref{fig:total_mass}. Figure~\ref{fig:total_mass}a shows the total stellar mass with a bin width of 0.1  membership probabilities ($\Delta P=0.1 $), whereas Figure~\ref{fig:total_mass}b represents the cumulative values of this distribution. The total mass of the stars with probabilities $P>0.9$ is 106$M_{\odot}$, while the total mass  reaches up to 162$M_{\odot}$ when the stars with probabilities $P>0$ are considered. Also, when considering the total mass of the stars with probabilities $P>0.5$, the total mass of the cluster is 117 $M/ M_{\odot}$, corresponding to 72 percent of the total mass of stars in all membership probabilities.

\section{Conclusion}
We presented photometric, astrometric, and kinematic analyses of open cluster Trumpler 2 using {\it Gaia} DR3 data. We identified 92 members as the most probable members with probabilities $P\geq 0.5$ in the direction of the cluster. These stars were used to estimate fundamental astrophysical and Galactic orbit parameters. We obtained age, distance modulus and reddening simultaneously from a {\it Gaia}-based CMD, whereas the metallicity is taken from the literature \citet{Carrera_2022}. The results are listed in Table~\ref{tab:Final_table}. The main outcomes of current study are summarized as follows:

\begin{table}
\renewcommand{\arraystretch}{1.2}
\setlength{\tabcolsep}{15pt}
  \centering
  \caption{ Fundamental parameters of Trumpler 2.}
  {\normalsize
        \begin{tabular}{lr}
\hline
Parameter & Value\\
\hline
($\alpha,~\delta)_{\rm J2000}$ (Sexagesimal)& 02:36:55.7, $+$55:54:18 \\
($l, b)_{\rm J2000}$ (Decimal)              & 137.3863, $-03.9778$    \\    
$f_{0}$ (stars arcmin$^{-2}$)               &  1.55$\pm$0.33          \\
$f_{\rm bg}$ (stars arcmin$^{-2}$)          & 12.59$\pm$0.02          \\
$r_{\rm c}$ (arcmin)                        & 2.48$\pm$1.09           \\
$r_{\rm lim}$ (arcmin)                      & 12                      \\
$r$ (pc)                                    & 2.39                    \\
Cluster members ($P\geq0.5$)                & 92                      \\
$\mu_{\alpha}\cos \delta$ (mas yr$^{-1}$)   & $1.494 \pm 0.004$       \\
$\mu_{\delta}$ (mas yr$^{-1}$)              & $-5.386 \pm 0.005$      \\
$\varpi$ (mas)                              & $1.455 \pm 0.059$       \\
$d_{\varpi}$ (pc)                           & $687\pm 27$             \\
$E(B-V)$ (mag)                              & $0.320\pm 0.013$ \\
$E(G_{\rm BP}-G_{\rm RP})$ (mag)            & $0.452\pm 0.019$ \\
$A_{\rm G}$ (mag)                           & $0.842\pm 0.035$ \\
$[{\rm Fe/H}]$ (dex)                        & $-0.262 \pm 0.106$$^{*}$ \\
Age (Myr)                                   & $110 \pm 10$            \\
Distance modulus (mag)                      & $10.027 \pm0.149$ \\
Isochrone distance (pc)                     & $686 \pm49$              \\
$(X, Y, Z)_{\odot}$ (pc)                    & ($-504$, 464, -48)      \\
$R_{\rm gc}$ (kpc)                          & 8.52                    \\
MF slope                                    & $1.33\pm 0.13$          \\
Total mass ($M/M_{\odot})$                  & 162                     \\
$V_{\gamma}$ (km s$^{-1}$)                  & $-4.48 \pm 1.11$        \\
$U_{\rm LSR}$ (km s$^{-1}$)                 & $5.12	\pm 1.34$         \\
$V_{\rm LSR}$ (kms$^{-1}$)                  & $2.11 \pm 0.35$         \\
$W_{\rm LSR}$ (kms$^{-1}$)                  & $-7.24\pm1.09$          \\
$S_{_{\rm LSR}}$ (kms$^{-1}$)               & $9.12	\pm1.76$          \\
$R_{\rm a}$ (pc)                            & $8752\pm 107$           \\
$R_{\rm p}$ (pc)                            & $8438 \pm 10$           \\
$z_{\rm max}$ (pc)                          & $101\pm 17$             \\
$e$                                         & $0.018\pm 0.007$        \\
$P_{\rm orb}$ (Myr)                         & $242 \pm 1$             \\
Birthplace (kpc)                            & $8.71 \pm 0.08$         \\
\hline
$^{*}$\citet{Carrera_2022}
        \end{tabular}%
    } 
    \label{tab:Final_table}%
\end{table}%

\begin{enumerate}
\item{We derived structural parameters from the RDP analyses. Analyses concluded that the background stellar density, central stellar density and the core radius values of the cluster are $f_{\rm bg}=12.59\pm0.02$ stars arcmin$^{-2}$, $f_0=1.55\pm0.33$ stars arcmin$^{-2}$ and $r_{\rm c}=2.48\pm1.09$ (arcmin), respectively. The limiting radius was determined by visual eye from RDP fitting. It was adopted as the point  where the RDP model matches with the background density level. Accordingly, the value of limiting radius was concluded to be $r_{\rm lim}^{\rm obs}=12$ arcmin.} 

\item{On the basis of the VPD, we estimated mean proper-motion components as ($\mu_{\alpha}\cos \delta, \mu_{\delta})=(1.494 \pm 0.004, -5.386 \pm 0.005$) mas yr$^{-1}$.}

\item{The metallicity value ($[{\rm Fe/H}]=-0.262 \pm 0.106$ dex) of the cluster was taken from the study of \citet{Carrera_2022} and transformed to  the mass fraction $z=0.0088$ to estimate age and distance modulus.}

\item{The {\it Gaia} based colour excess was estimated as $E(G_{\rm BP}-G_{\rm RP})=0.452\pm 0.019$} mag by comparing the cluster CMD with the theoretical isochrone (\citet{Bressan_2012} with $z=0.0088$. This value corresponds to a $U\!BV$-based colour excess to be $E(B-V)=0.320\pm 0.013$ mag via the use of expression $E(G_{\rm BP}-G_{\rm RP})=1.41\times E(B-V)$ as given by \citet{Sun_2021}.

\item{The isochrone fitting distance to the open cluster Trumpler 2 was derived as $d_{\rm iso}=686\pm 49$ pc. This value is matched well by the distance obtained from the mean trigonometric parallax $d_{\varpi}$= $687\pm 27$ pc. The age was evaluated as $t=110 \pm 10$ Myr by fitting the isochrone with $z=0.0088$ scaled given by \citet{Bressan_2012} to the observable CMD.}

\item The mass function slope of Trumpler 2 was obtained as $\Gamma=1.33\pm 0.13$. This slope is in good agreement with the value of 1.35 given by \citet{Salpeter_1955}. The total mass of Trumpler 2 is about 162 $M/M_{\odot}$.

\item{Galactic orbit analyses showed that Trumpler 2 is orbiting in a boxy pattern outside the solar circle and belongs to young thin-disc component of the Milky Way. Moreover, the birth-radius ($8.71\pm 0.08$ kpc) indicated that the cluster was formed outside the solar circle.}

\end{enumerate}

\section*{Acknowledgements}
 We are grateful to the anonymous referees for their feedback, which improved the paper. This study has been supported in part by the Scientific and Technological Research Council (T\"UB\.ITAK) 122F109. This study is a part of the master thesis of Seval Ta\c {s}demir. This research has made use of the WEBDA database, operated at the Department of Theoretical Physics and Astrophysics of the Masaryk University. We also  made use of NASA's Astrophysics Data System as well as the VizieR and Simbad databases at CDS, Strasbourg, France and data from the European Space Agency (ESA) mission \emph{Gaia}\footnote{https://www.cosmos.esa.int/gaia}, processed by the \emph{Gaia} Data Processing and Analysis Consortium (DPAC)\footnote{https://www.cosmos.esa.int/web/gaia/dpac/consortium}. Funding for DPAC has been provided by national institutions, in particular the institutions participating in the \emph{Gaia} Multilateral Agreement.



\bibliographystyle{mnras}
\bibliography{refs}

\bsp	
\label{lastpage}
\end{document}